\documentclass[longauth]{aa}  

\usepackage{graphicx}
\usepackage{amsmath}
\usepackage{natbib}
\usepackage{txfonts}
\usepackage{gensymb}
\usepackage{threeparttable}
\usepackage{booktabs,caption}
\usepackage{textcomp}
\usepackage[normalem]{ulem}
\usepackage{hyperref}
\usepackage{siunitx}
\hypersetup{colorlinks=true,linkcolor=blue,citecolor=blue,urlcolor=blue}
\newcommand\brg{Br$\gamma$}
\newcommand\umdot{M${_\odot}\,yr^{-1}$}
\newcommand\rco{$R_{\mathrm{cor}}$}
\newcommand\rsub{$R_{\mathrm{sub}}$}
\newcommand\rtr{$R_{\mathrm{tr}}$}

\usepackage[dvipsnames]{xcolor}

\newcommand\AS[1]{{\color{purple} #1}} 

\begin{document} 

   \title{The GRAVITY Young Stellar Object survey\thanks{Based on observations collected at the European Organisation for Astronomical Research in the Southern Hemisphere under ESO programs 106.212G.004 and 108.228Z.005}}
   
   \subtitle{XI. Probing the inner disk and magnetospheric accretion region of CI Tau}
   
   \author{GRAVITY Collaboration: A. Soulain
          \inst{1}
          \and
          K. Perraut
          \inst{1}
          \and
          J. Bouvier
          \inst{1}
          \and
          G. Pantolmos
          \inst{1}
          \and
          A. Caratti o Garatti
          \inst{4}
          \and
          P. Caselli
          \inst{5}
          \and
          P. Garcia
          \inst{6}
          \and
          R. Garcia Lopez
          \inst{4}
          \and
          N. Aimar
          \inst{9}
          \and 
          A. Amorin
          \inst{6, 7}
          \and 
          M. Benisty
          \inst{1}
          \and 
          J.-P. Berger
          \inst{1}
          \and 
          G. Bourdarot
          \inst{5}
          \and 
          W. Brandner
          \inst{8}
          \and
          Y. Clénet
          \inst{9}
          \and 
          T. de Zeeuw
          \and 
          R. Davies
          \inst{5}
          \and 
          A. Drescher
          \inst{5}
          \and 
          A. Eckart 
          \inst{2, 3}
          \and 
          F. Eisenhauer
          \inst{5}
          \and 
          N.M. Förster Schreiber
          \inst{5}
          \and 
          E. Gendron
          \inst{9}
          \and
          R. Genzuel
          \inst{5}
          \and
          S. Gillessen
          \inst{5}
          \and
          G. Hei\ss el
          \inst{9, 10}
          \and
          Th. Henning
          \inst{8}
          \and
          S. Hippler
          \inst{8}
          \and
          M. Horrobin
          \inst{11}
          \and
          L. Jocou
          \inst{1}
          \and
          P. Kervella
          \inst{9}
          \and
          L. Labadie
          \inst{2}
          \and
          S. Lacour
          \inst{9}
          \and
          V. Lapeyrère
          \inst{9}
          \and
          J.-B. Le Bouquin
          \inst{1}
          \and
          P. Léna
          \inst{9}
          \and
          D. Lutz
          \inst{5}
          \and
          F. Mang
          \inst{5}
          \and
          T. Ott
          \inst{5}
          \and
          T. Paumard
          \inst{9}
          \and 
          G. Perrin
          \inst{9}
          \and
          J. Sanchez
          \inst{14, 8}
          \and 
          S. Scheithauer
          \inst{8}
          \and
          J. Shangguan
          \inst{5}
          \and
          T. Shimizu
          \inst{5}
          \and
          O. Straub
          \inst{5, 12}
          \and
          C. Straubmeier
          \inst{2}
          E. Sturm
          \inst{5}
          \and
          L.J. Tacconi
          \inst{5}
          \and
          F. Vincent
          \inst{9}
          \and
          E. van Dishoeck
          \inst{5, 15}
          \and
          F. Widmann
          \inst{5}
          \and
          E. Wieprecht
          \inst{5}
          \and
          E. Wiezorrek
          \inst{5}
          \and
          S. Yazici
          \inst{5}
          }

    \institute{Univ. Grenoble Alpes, CNRS, IPAG, 38100 Grenoble, France 
        \and I. Physikalisches Institut, Universität zu Köln, Zülpicher Str. 77, 50937, Köln, Germany 
        \and Max-Planck-Institute for Radio Astronomy, Auf dem Hügel 69, 53121 Bonn, Germany 
        \and INAF-Osservatorio Astronomico di Capodimonte, Salita Moiariello 16, I-80131 Napoli, Italy 
        \and Max Planck Institute for Extraterrestrial Physics, Giessenbachstr. 1, D-85748 Garching, Germany 
        \and CENTRA - Centro de Astrofísica e Gravitação, IST, Universidade de Lisboa, 1049-001 Lisboa, Portugal 
        \and Universidade de Lisboa - Faculdade de Ciências, Campo Grande, 1749-016 Lisboa, Portugal 
        \and Max Planck Institute for Astronomy, Königstuhl 17, 69117 Heidelberg, Germany 
        \and LESIA, Observatoire de Paris, Université PSL, CNRS, Sorbonne Université, Université de Paris, 5 place Jules Janssen, 92195 Meudon, France 
        \and Advanced Concepts Team, European Space Agency, TEC-SF, ES-TEC, Keplerlaan 1, 2201, AZ Noordwijk, The Netherlands 
        \and 1st Institute of Physics, University of Cologne, Zülpicher Straße 77, 50937 Cologne, Germany 
        \and ORIGINS Excellence Cluster, Boltzmannstraße 2, D-85748 Garching, Germany 
        \and Max Planck Institute for Radio Astronomy, auf dem Hügel 69, D-53121 Bonn, Germany 
        \and Instituto de Astronomía, Universidad Nacional Autónoma de México, Apdo. Postal 70264, Ciudad de México, 04510, México 
        \and Sterrewacht Leiden, Leiden University, Postbus 9513, 2300 RA, Leiden, The Netherlands 
       }

   \date{\today}
 
  \abstract
   {T Tauri stars are known to be the cradle of planet formation. Most exoplanets discovered to date lie at the very inner part of the circumstellar disk (< 1 au). The innermost scale of Young Stellar Objects is therefore a compelling region to be addressed, and long-baseline interferometry is a key technique to unveil their mysteries.}
   {We aim at spatially and spectrally resolving the innermost scale ($\leq 1\,\mathrm{au}$) of the young stellar system CI Tau to constrain the inner disk properties and better understand the magnetospheric accretion phenomenon.}
   {The high sensitivity offered by the combination of the four 8-m class telescopes of the Very Large Telescope Interferometer (VLTI) allied with the high spectral resolution (R~$\sim$~4000) of the K-band beam combiner GRAVITY offers a unique capability to probe the sub-au scale of the CI Tau system, tracing both dust (continuum) and gas (Br$\gamma$ line) emission regions. We develop a physically motivated geometrical model to fit the interferometric observables (visibilities and closure phases (CP)) and constrain the physical properties of the inner dusty disk. The continuum-corrected pure line visibilities have been used to estimate the size of the Hydrogen I \brg{} emitting region.} 
   {From the K-band continuum study, we report an highly inclined ($i\sim70\degree$) resolved inner dusty disk, with an inner edge located at a distance of $21\pm2\,R_\star$ from the central star, which is significantly larger than the dust sublimation radius (\rsub{}$= 4.3$ to $8.6\,R_\star$). The inner disk appears misaligned compared to the outer disk observed by ALMA and the non-zero closure phase indicates the presence of an asymmetry that could be reproduced with an azimuthally modulated ring with a brighter south-west side.. From the differential visibilities across the \brg{} line, we resolve the line emitting region, and measure a size of $4.8^{+0.8}_{-1.0}$ $R_\star$.}
   {The extended inner disk edge compared to the dust sublimation radius is consistent with the claim of an inner planet, CI Tau b, orbiting close-in.
   The inner-outer disk misalignment may be induced by gravitational torques or magnetic warping. The size of the \brg{} emitting region is consistent with the magnetospheric accretion process. Assuming it corresponds to the magnetospheric radius, it is significantly smaller than the co-rotation radius (\rco{}$=8.8\pm1.3\,R_\star$), which suggests an unstable accretion regime that is consistent with CI Tau being a burster.}


\maketitle
%
\section{Introduction}

The power of long baseline near-infrared interferometry to investigate the inner regions of young stellar systems has been amply demonstrated in the past years \citep{2010ARA&A..48..205D}. The inner disk structure \citep{Gravity21}, associated outflows \citep{Gravity17}, and the accretion process \citep{Gravity20} can all be probed on an angular scale of less than one millisecond of arc (mas), which corresponds to a region extending a few stellar radii around the central star at the distance of the closest star-forming regions. On this scale, accretion in classical T Tauri stars (i.e., Class II young stellar objects with $M_{\star}$ < 2 $M_{\odot}$) occurs along funnel flows due to the strong stellar magnetic field ($\approx$ kG) that channels the infalling gas \citep[e.g.,][]{Romanova15, Hartmann16, Bouvier07}. The inner disk is disrupted at the magnetospheric or truncation radius (typically at $\sim5$ $R_{\star}$), where the magnetic pressure of the stellar field balances the thermal and/or ram pressure of the accreting matter \citep{Bessolaz08, blinova16, Pantolmos20}.


The observational evidence for the magnetospheric accretion process in young stars, while quite convincing and widely accepted, has so far been mostly indirect. It relies on measurements of magnetic field strength and topology \citep[e.g.,][]{Donati09} and mass accretion rate estimates \citep[e.g.,][]{Manara21, Alcala21}. It is probed through a number of spectral diagnostics, including the emission line spectrum of T Tauri stars that forms, at least in part, in the magnetic funnel flows \citep[e.g.,][]{Bouvier20a}, and the UV continuum excess arising for the accretion shock at the stellar surface \citep[e.g.,][]{Espaillat22}. In recent years, the increased sensitivity of long baseline interferometers has opened a new window to the star-disk interaction region, with results that provide a direct estimate of the extent of the magnetospheric cavity and support the magnetospheric accretion paradigm \citep{Gravity20, 2020A&A...636A.108B, Wojtczak23}.

\begin{table*}[ht!]
    \centering
            \caption{\label{tab:log} Journal of the VLTI/GRAVITY observations.}
    	\renewcommand{\arraystretch}{1.3}
    		\begin{tabular}{l l l l l l l l }
    		\hline
    		\hline
    		MJD & Date & Time (UT) & Configuration & $N$ & Seeing ('') & $\tau_0$ (ms) & Calibrators\\
            \hline
            59223.12 & 2021-01-09 & 01:35--04:03 & UT1-UT3-UT4 & 11 & 0.68--1.05 & 2.7--5.7 & HD~31464, HD~40003\\
            59633.04 & 2022-02-23 & 00:39--01:26 & UT1-UT2-UT3-UT4 & 6 & 0.36--0.53 & 5.8--8.1 & HD~31464, HD~40003\\
            \hline
        \end{tabular}
     {\\\raggedright \textbf{Notes.} $N$ denotes the number of calibrated points recorded on the target. \par}
\end{table*}

 We present here the results from VLTI/GRAVITY observations of the young stellar system CI Tau. CI Tau is a 2 Myr-old \citep{Guilloteau14}, 0.9 $M_\odot$ \citep{Simon19} classical T Tauri star, located at a distance of 160.3 $\pm$ 0.4 pc \citep{Gaia22} in the Taurus molecular cloud. It is known to harbour a strong, mostly poloidal magnetic field up to 3.7 kG and exhibits a variable mass accretion rate of the order of 2$\times$10$^{-8}\,M_\odot\,$yr$^{-1}$ \citep{2020MNRAS.491.5660D}. On the large scale, CI Tau is surrounded by a circumstellar disk that extends up to 200 au on millimetre continuum images, and features a succession of dusty rings, with gaps located at radii $\sim$ 13, 39, and 100 au, suggestive of on-going planet formation \citep{2018ApJ...866L...6C}. Indeed, CI Tau is the only accreting T Tauri star for which a hot super-Jupiter ($M_\mathrm{p}$ = 11.3 $M_{\mathrm{Jup}}$) has been claimed from radial velocity variations \citep{2016ApJ...826..206J}, although the planetary origin of the radial velocity signal has been questioned \citep{2020MNRAS.491.5660D}. 

The goal of the VLTI/GRAVITY observations we report here was to investigate the star-disk interaction region of this intriguing young system, to derive the properties of the dusty inner disk on a scale of 0.1 au or less from continuum K-band visibilities and phases, and to investigate the magnetospheric accretion region through the analysis of differential interferometric quantities measured across the Br$\gamma$ line profile. Section 2 describes the observations and data reduction, Section 3 presents the derivation of the properties of the inner disk and of the Br$\gamma$-line emitting region through model-fitting, and Section 4 discusses the results in light of the possible existence of CI Tau b, compares the inner disk properties to the outer disk structure, and confront the interferometric results to magnetospheric accretion models. Conclusions are presented in Section 5. 

\section{Observations}

We observed CI Tau at two epochs on January 9$^\mathrm{{th}}$ 2021 and February 23$^\mathrm{{rd}}$ 2022 in the K-band with the GRAVITY instrument \citep{2017A&A...602A..94G}, combining the four Unit Telescopes (UTs) of the ESO Very Large Telescope Interferometer (VLTI) installed in Paranal, Chile. This program was part of the GTO large program dedicated to the Young Stellar Objects (YSO). The maximum baseline accessible with the UTs is 130 m, which corresponds to a maximal angular resolution of $\lambda/2B_{max} \approx 1.5$ mas at 2.2~$\mu$m. Both epochs were carried out using the single-field on-axis mode, where 50$\%$ of the flux is sent to the fringe tracker (FT) and 50$\%$ to the scientific instrument (SC): the instrument tracks the fringes on the science target itself to stabilize them at a frequency of 900 Hz \citep{2019A&A...624A..99L}, enabling longer integration on the SC, in particular for faint targets. Data were obtained in high spectral resolution mode (R~$\sim$~4000). GRAVITY covers a spectral range from 1.9 to 2.4 $\mu$m, including the neutral-hydrogen \brg{} line at 2.1661 µm. Weather conditions were excellent during the two nights; we recorded eleven and six 5-min long files on the object in 2021 and 2022, respectively (Table \ref{tab:log}). We observed two calibrators before (HD~31464) and after (HD~40003) the observations to accurately estimate the atmospheric transfer function and calibrate the interferometric observables. We used SearchCal tool \citep{2016A&A...589A.112C} to establish our calibrator list, which offers a way to search for objects that are single stars, bright, unresolved and close to the target. Due to technical issues during the first epoch, one of the telescopes (UT2) was down during the observations, which reduced the number of exploitable baselines from six to three.

The data reduction was performed using the ESO GRAVITY pipeline\footnote{\url{https://www.eso.org/sci/software/pipelines/gravity.}}\citep{2014SPIE.9146E..2DL}. For each file, we extracted six (three) complex visibilities and four (one) closure phase measurements in 2022 (2021), dispersed over six spectral channels for the FT and about 1600 for the SC, respectively. The bluest part of the fringe tracker being contaminated by the metrology laser working at 1.908 $\mu$m, we discarded the first channel from our analysis. Finally, we recovered the differential visibilities and phases in the \brg{} line region from the SC data. The error bars supplied by the pipeline are known to be underestimated and do not include residual calibration effects \citep{2020A&A...636A.108B, 2021A&A...655A..73G}. To be conservative, we refined our uncertainties by computing the total rms over the files for both observables, which yields constant uncertainties of 2$\%$ for the visibility and 0.7 degrees for the closure phases. The final uncertainties being similar between the two epochs, we adopted the same error bars for all observations. Normalising uncertainties between our two epochs allows us to mitigate the effects of different weather conditions and adaptive optics correction, and to attribute the same weight to the 2021 and 2022 data sets.

\section{Results}

In this section we report  the method used to derive the main properties of the emitting regions both in the K-band continuum and across the \brg{} line.

\subsection{The inner dusty disk}

\subsubsection{Geometrical model}

To model the continuum complex visibility, we follow the same approach as adopted by \citet{2017A&A...599A..85L} and \citet{2021A&A...655A..73G}, which consists of representing the system as a three-component model: an unresolved point-like star ($s$) as we are not able to resolve the stellar photosphere, a circumstellar dusty disk ($d$), and a fully resolved component ($h$). Each element is represented by a complex visibility function ($V_s$, $V_d$ and $V_h$) and accounted for in the whole system by their flux contributions ($F_s$, $F_d$ and $F_h$):

\begin{equation}
    V_{tot}(\Vec{B}/\lambda) =  \frac{F_s V_s + F_d V_d(\Vec{B}/\lambda) + F_h V_h}{F_s+F_d+F_h},
\end{equation}
where $V_s = 1$ for a point-source, $V_h = 0$ for a fully resolved component, $\Vec{B}/\lambda$ is the spatial frequency in rad$^{-1}$ at the different baselines $\Vec{B}$ and $F_s+F_d+F_h = 1$. We consider wavelength independent flux contributions (non-chromatic model). The extended component $V_h$ is commonly used to mimic the effect of the scattered light \citep{2008ApJ...673L..63P}, which decreases the visibility at the zero spatial frequency ($\Vec{B}/\lambda=0$). This last component appears to contribute significantly in the case of YSO, such as transitional disks \citep{2017A&A...599A..85L} or T Tauri stars \citep{2015A&A...574A..41A, 2021A&A...655A..73G}. As \citet{2017A&A...599A..85L}, we model the dusty disk contribution by an circular ring defined by a radius $a_r$, an inclination $i$, and a position angle $PA$. To describe a smooth inner rim radial profile, we convolve the ring model by a 2-d Gaussian model. In the following, we present the convolution effect by using the ratio between the Gaussian kernel $a_k$ and the half-flux radius $a$ as $w=a_k/a$ in percent.

Finally, we add a brightness azimuthal modulation along the ring described by cosine and sine amplitudes $c_1$ and $s_1$. This modulation can be used to represent a non-uniform azimuthal disk profile responsible for a non-zero closure phase signature. In practice, $c_1$ and $s_1$ can vary between -1 and 1, allowing us to drag the brightest portion (if any) around the disk in polar coordinates.
\subsubsection{Fitting strategy}

For the first epoch, the FT data were not fully exploitable due to a relatively low coherence time ($\sim$ 2-3 ms) that degraded the signal-to-noise ratio significantly. To address this, we used the SC data instead and calculated the observables averaged over 300 spectral channels, which reproduces the spectral resolution of the FT camera (R~$\sim$~30). For the second epoch, the weather conditions were optimal with a coherence time around 7 ms, but we adopted the same approach as in 2021 to get consistent results between the two epochs.

To estimate the properties of the continuum emitting region, we perform the fit over several steps to avoid any local $\chi^2$ minima and robustly estimating the associated uncertainties. Since the circumstellar disk is only partially resolved by the interferometer ($V\sim0.8$), its flux contribution $F_d$ and its size $a_r$ are partly degenerated \citep{2017A&A...599A..85L}. To get an independent estimate of the relative contributions of the disk and the star in the K-band, we used the near-infrared veiling measured as described in \citet{Sousa23}. At the time of our 2021 observations, the infrared veiling amounted to $r_K = 0.83\pm0.04$ (A. Sousa, priv. comm.), which yields an estimate of $F_s = 1/(1+r_K) = 55\%$ around 2.2 $\mu$m. To consider the star's intrinsic variability, we adopted a typical error of 5\% on this measurement. Besides, we independently evaluated the stellar contribution by fitting the target's spectral energy distribution. We collected the photometry measurements from EPIC \citep[B, V and R bands, ][]{2014PASP..126..398H, 2017yCat.4034....0H}, Gaia DR3 \citep[G$_{bp}$, G and G$_{rp}$ bands, ][]{Gaia22} and 2MASS \citep[J, H, and K bands, ][]{2003yCat.7233....0S, 2006AJ....131.1163S}. We adopted the stellar parameters and the visual extinction ($A_V = 0.65$) determined by \citet{2020MNRAS.491.5660D} and use the accurate distance estimate from Gaia DR3 \citep[$160.3\pm0.4$ pc,][]{Gaia22}. We thus derived $F_s = 55\%$, quite consistent with the veiling measurement. We therefore used this value for the star contribution as a prior during the fitting process, with a 5$\%$ tolerance. This additional constraint releases the degeneracy between the ring's size and its flux contribution.

We carried out an initial parameter search using the Levenberg-Marquardt method\footnote{Available with scipy \citep{scipy}}. We estimated the geometrical parameters with and without azimuthal modulation by using or not the closure phase quantity. Given the lower $\chi^2$ values obtained with the asymmetric case (1.6 versus 2.1 for the total $\chi^2$ value, and 0.5 versus 2.9 when considering the CP only), we adopted this model to fit the data using a Monte-Carlo Markov Chain (MCMC) approach\footnote{Available with \texttt{emcee} \citep{2013PASP..125..306F}.}. We used 200 walkers for 2000 iterations and rejected the first 1000 iterations as the burn-in time. The 1-$\sigma$ uncertainty associated with each parameter is computed from the final distribution of walkers using the 16, 50, 84$\%$ percentiles.

\begin{figure}
    \centering
    \includegraphics[width=0.9\columnwidth]{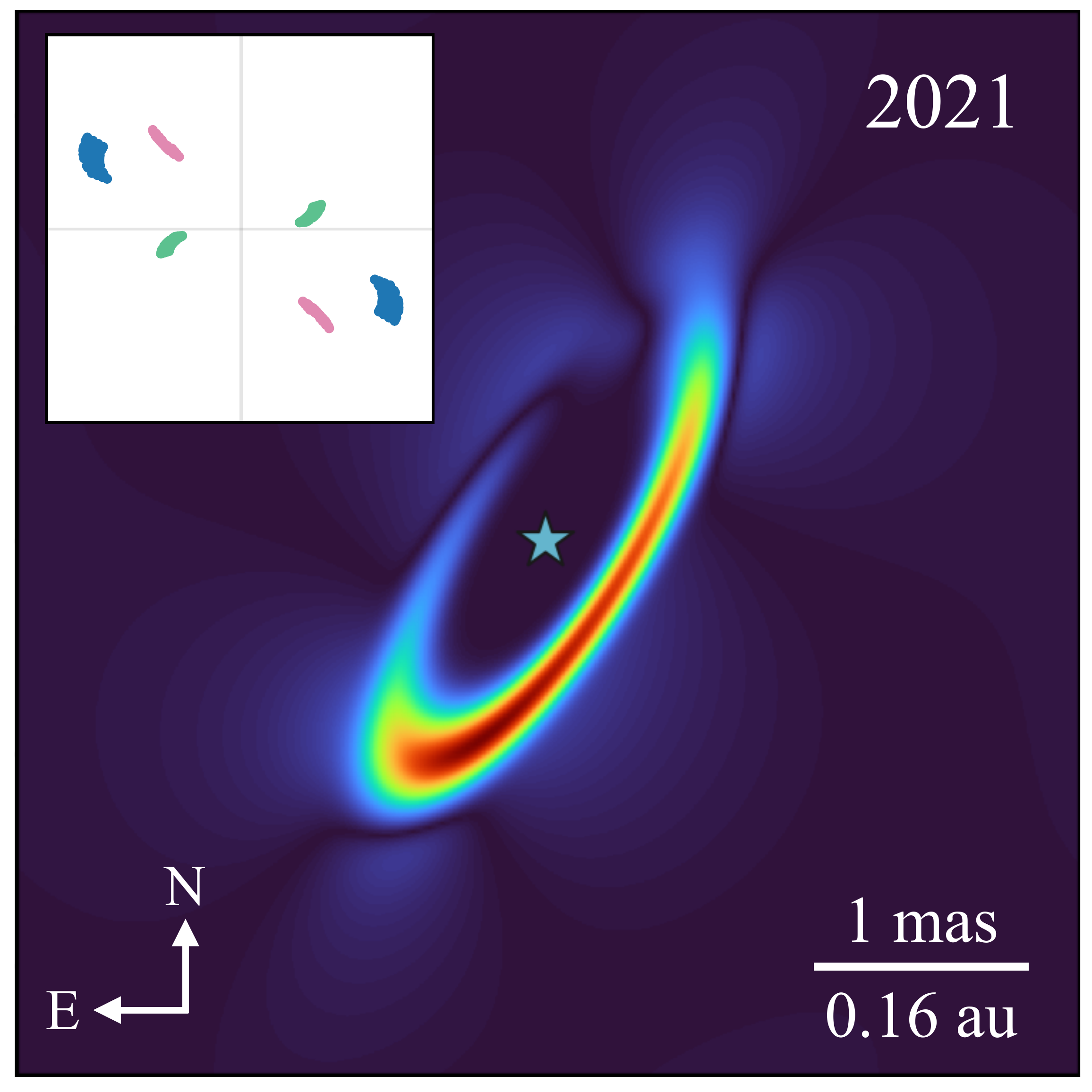}
    \caption{\label{fig:model_image} Model image of CI Tau's inner disk in 2021. The position of the star is depicted and has been removed to show the disk structure. The upper-left inset shows the u-v coverage. The colour circles represent the baselines: UT1-UT3 (\textit{pink}), UT3-UT4 (\textit{green}) and UT1-UT4 (\textit{blue}).}
\end{figure}

\subsubsection{Inner disk properties}

For the 2021 data set, the model converges toward an elongated thin ring model with an inner rim radius of $a_r = 0.20\pm0.02$~au. We estimate a width-to-radius ratio $w$ smaller than $28\%$, indicating a resolved inner gap. The major-to-minor axis elongation corresponds to a relatively high inclination of $i=71\pm1\degree$ at a position angle of $PA=148\pm1\degree$ counted from North to East. The dusty disk contribution is constant between the two epochs ($F_d = 36 \pm 2\%$) and the halo contribution remains between 8 and 10$\%$. For the second epoch, the limited time of observation ($\sim$1h) corresponds accordingly to a short range of spatial frequencies (30 vs. 75 arcsec$^{-1}$, see Fig. \ref{fig:fit_V2_2021} and \ref{fig:fit_V2_2022}). This prevents us from resolving the inner gap ($w$ close to 1), and from constraining the orientation of the system in an unambiguous way. In order to derive the lower limit of the system's inclination for the second epoch, we performed a $\chi^2$-minimum search (see App. \ref{app:chi2_fit_incl} for details). The inner dusty disk properties are presented in Table \ref{tab:results_fit_cont}. The values of inclination and position angle for the second epoch correspond to those obtained from the $\chi^2$ search (Fig. \ref{fig:chi2_orient_limit}). The MCMC-posterior distribution obtained for the second epoch converges to a very high inclination (close to $90\degree$, Fig. \ref{fig:mcmc_distribution2}) that prevents us from determining the asymmetric modulation ($c_j$, $s_j$ compatible with zero). The inner rim radius estimate from the 2022 data set appears significantly smaller than the one derived for the 2021 data set. The inner gap being unresolved in 2022, the inner disk size could be underestimated.

Figure \ref{fig:model_image} displays the best-fit model image as determined by GRAVITY in 2021. The non-zero closure phases are consistent with the presence of an asymmetry in the inner rim located in the South-West part. The data-model comparison and the MCMC distributions are presented and discussed in Appendix \ref{app:fit_results} and \ref{app:mcmc_distr}.
\begin{table}[ht]
    \centering
            \caption{\label{tab:results_fit_cont} Best-fit parameters of the K-band continuum VLTI/GRAVITY data of CI Tau obtained in 2021 and 2022 with $1\sigma$ error bars. }
    	\renewcommand{\arraystretch}{1.3}
    		\begin{tabular}{l l l}
    		\hline
    		\hline
    		Parameters & 2021 & 2022\\
            $F_d$ [$\%$] &  $36\pm2$ & $35 \pm 5$\\
            $F_h$ [$\%$] &  $9.2\pm0.6$ & $8.3 \pm 0.2$\\
            $i$ [$\degree$] &  $71 \pm 1$ & $\geq 70$\\
            $PA$ [$\degree$] &  $148 \pm 1$ & $140^{+16}_{-12}$\\
            $c_1 $ &  $0.94^{+0.04}_{-0.08}$ & -\\
            $s_1 $ &  $-0.75^{+0.09}_{-0.12}$ & -\\
            $w$ [$\%$] & 17$^{+11}_{-6}$ & unresolved\\
            $a_r$ [mas] &  $1.25 \pm 0.13$ & $0.81 \pm 0.13$\\
            $a_r$ [au]$^1$ &  $0.20 \pm 0.02$ & $0.13 \pm 0.02$\\
            $a_r$ [R$_{\star}$]$^2$ &  $21 \pm 2$ & $14\pm 2$\\
            $\chi^2_{r}$ & 1.56 & 0.87\\
            \hline
            \multicolumn{3}{l}{$^1$ We used the Gaia distance $160.3\pm0.4$ pc}\\
            \multicolumn{3}{l}{$^2$ We adopted a stellar radius of 2$\,R_\odot$ \AS{\citep{2020MNRAS.491.5660D}}.}\\
            
        \end{tabular}
\end{table}

\subsection{The Br$\gamma$ line emitting region}

GRAVITY's high spectral resolution allows us to resolve the \brg{} line profile at 2.1661 µm. This spectral feature is the privileged tracer of the star-disk interaction, attributed to the magnetospheric accretion process \citep{1994ApJ...426..669H}. Following \citet{2007A&A...464...87W}, \citet{2008A&A...489.1157K} and \citet{Wojtczak23}, we compute the continuum-subtracted observables, the so-called pure line visibilities, by using the emission line profile provided by GRAVITY. This differential observable is only sensitive to the \brg{} emitting region and remove all contributions from the star and disk, assuming no photospheric absorption is present in the line region, which is adequate for cooler T Tauri stars. The pure line visibility $V_{line}(\lambda)$ is computed as: 

\begin{equation}
    \label{eq:vpure}
    V_{Line}(\lambda) = \frac{F_{L/C}(\lambda)V_{Tot}(\lambda) - V_{Cont}}{F_{L/C}(\lambda)-1},
\end{equation}

\begin{figure}
    \centering
    \includegraphics[width=0.98\columnwidth]{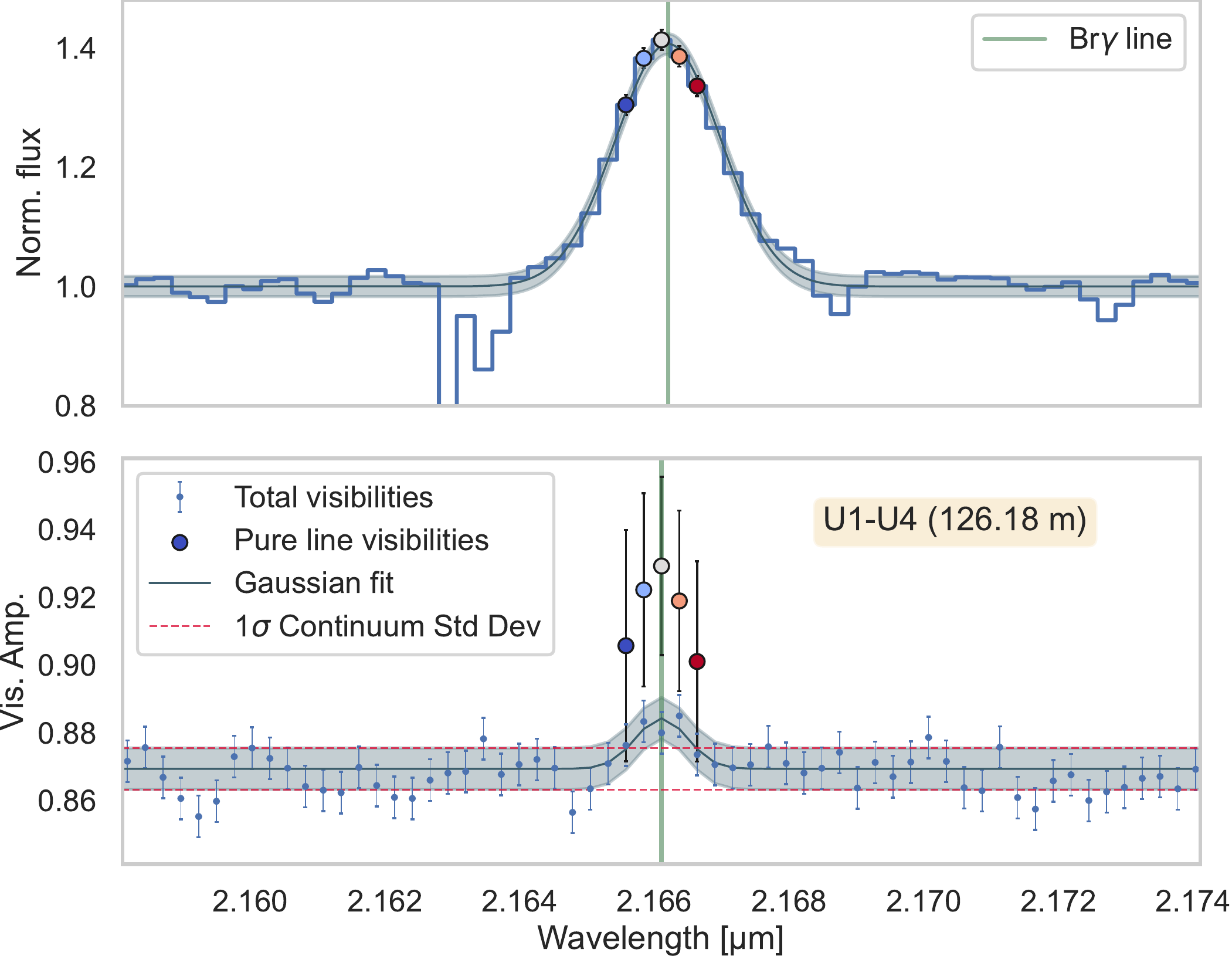}
    \caption{\label{fig:diff_vis} \brg{} line observables. \textbf{Top:} The normalized spectral line profile averaged over the four telescopes with GRAVITY. \textbf{Bottom:} Differential visibilities from the UT1-UT4 baseline of the CI Tau observation in 2021. The small blue dots and error bars represent the total visibility. The larger coloured dots indicate the pure line visibilities after the subtraction of the continuum contribution (see Eq. \ref{eq:vpure}). The continuum estimate and the associated uncertainty are shown as red dashed lines. The Gaussian model used to fit the total visibility is represented as a blue line.}  
\end{figure}

$F_{L/C}$ denotes the total line-to-continuum flux ratio as taken from the normalised spectrum (Fig.~\ref{fig:diff_vis}), $V_{Cont}$ is the visibility computed in the continuum, and $V_{Tot}$ is the total complex quantities measured by GRAVITY. 

In order to enhance the signal across the \brg{} line, we combine the 11 files available for the first epoch in 2021. The u-v plane rotation occurring during the observational sequence remained relatively small (< 10 degrees) and thus the files can be combined without degrading the scientific signal significantly. Unfortunately, the data quality in 2022 were not sufficient to reach the required signal-to-noise ratio to detect the differential signal. Figure \ref{fig:diff_vis} presents the \brg{} emission line profile, the total differential visibility and the extracted pure-line visibility. A significant signal is only detected for the most extended baseline (UT1-UT4, 126.16 m) with a 3-$\sigma$ detection in the visibility amplitude. We did not detect any significant differential phase signals neither epochs or baselines.

\begin{figure}[h!]
    \centering
    \includegraphics[width=0.98\columnwidth]{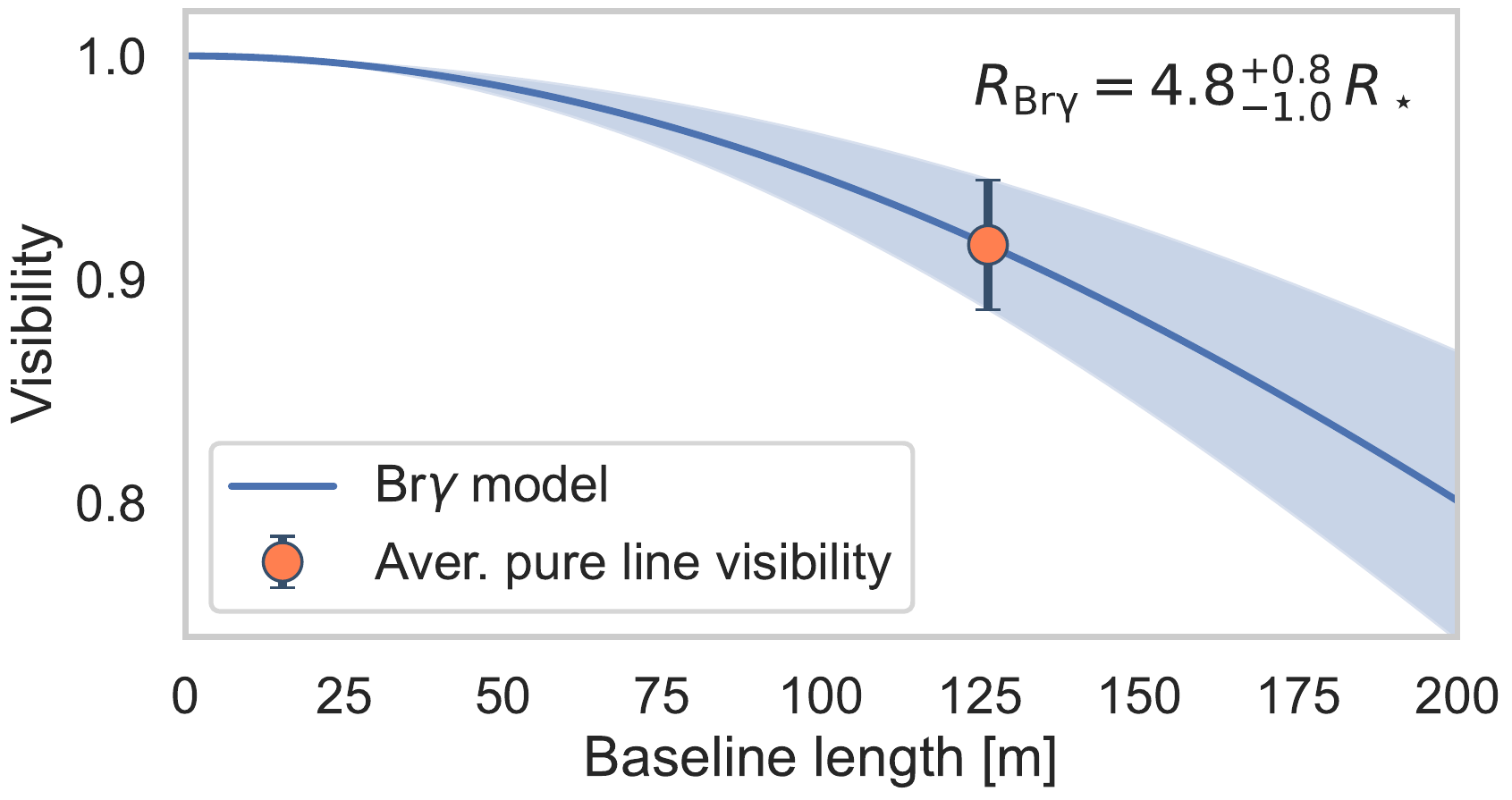}
    \caption{\label{fig:fit_size_mag} Comparison between the observed \brg{} visibility (orange dot) and a visibility curve predicted for a Gaussian disk model of the emitting region (blue curve). The blue shaded area depicts the uncertainty on the size relative to the visibility error.}
\end{figure}

The pure line visibilities across the \brg{} line profile range from 0.90 to 0.93, indicating a more compact emitting region than the inner disk seen in the continuum. In order to estimate the characteristic size of the \brg{} emitting region, we averaged the five pure line visibilities over the spectral channels and derived a unique visibility measurement of $V_{\mathrm{Br\gamma}}=0.92\pm0.03$. Based on a simple geometric Gaussian disk model \citep{2007NewAR..51..576B}, we extracted the half-flux radius (or half width half maximum, HWHM) corresponding to $V_{\mathrm{Br\gamma}}$. Figure \ref{fig:fit_size_mag} presents the visibility curve of a 2-d Gaussian model compared to the extracted pure line visibility. The visibility uncertainty of 0.03 is directly reported on the visibility curve model (blue shade area), which yields asymmetric errors on the half-flux radius estimate. We thus derive a \brg{} emission region radius of $R_{\mathrm{Br\gamma}} = 0.28^{+0.05}_{-0.06}$ mas, which corresponds to $0.045^{+0.008}_{-0.009}$~au at the distance of CI Tau, or $4.8^{+0.8}_{-1.0}$ $R_\star$ for a stellar radius of 2 $R_\odot$.  The \brg{} emitting region is thus significantly more compact than the continuum disk radius.

\subsection{Mass-accretion rate and truncation radius}
\label{sec:mass_acc_measure}

To estimate the instantaneous mass-accretion rate at the time of GRAVITY observations, we computed the \brg{} line luminosity and used the line-to-accretion luminosity relations from \citet{Alcala17}. We measured the equivalent width of the \brg{} line on the GRAVITY spectrum, EW$_{\mathrm{Br\gamma}}$, and estimated the extinction-corrected nearby continuum flux from the 2MASS K-band magnitude \citep{2006AJ....131.1163S}. With EW$_{\mathrm{Br\gamma}}=7.9\pm0.4\,\AA$ and a continuum flux of $3.3\,10^{-13}$ $\mathrm{W\,m^{-2}\,\mu m^{-1}}$, we derive a line luminosity of $(2.07\pm0.10\,)10^{-4}$ $L_\odot$ at $160.3\pm0.4$ pc. 

The accretion luminosity can then be derived from the empirical relationship \citep{Alcala17}:
\begin{equation}
\log\left(\frac{L_{acc}}{L_\odot}\right) = a \log\left(\frac{L_{Line}}{L_\odot}\right) + b,
\end{equation}
with $a=1.19\pm0.10$ and $b=4.02\pm0.51$. Finally, the accretion luminosity can be converted into an instantaneous mass-accretion rate using the following relation \citep{hartmann98}:
\begin{equation}
\label{eq:macc}
\dot{M}_{acc} = \left(1 - \frac{R_\star}{R_{\mathrm{Br\gamma}}}\right)^{-1}L_{acc}\frac{R_\star}{GM_\star},
\end{equation}
which assumes that the energy released by the infalling material confined within the magnetosphere is entirely converted into accretion luminosity. Adopting the GRAVITY size of the \brg{} emitting region $R_{\mathrm{Br\gamma}}$ $=$ 4.8$\,R_\star$ for the magnetosphere radius, we derive a mass accretion rate of $\dot{M}_{acc} = 3.9^{+12.8}_{-3.0}\,10^{-8}$~$M_\odot\,\mathrm{yr}^{-1}$ ($\log(\dot{M}_{acc}) = -7.4\pm0.6$).

The size of the magnetospheric accretion region, characterised by the magnetic truncation radius \rtr{}, is driven by the strength of the magnetic field and the mass accretion rate \citep{Hartmann16}:
\begin{equation}
    \label{eq:rtrunc}
    \frac{R_{\mathrm{tr}}}{R_\odot} = 12.6\frac{B^{4/7}R_2^{12/7}}{M_{0.5}^{1/7}\dot{M}_{-8}^{2/7}},
\end{equation}
where B is the surface field strength of the dipolar magnetic field at the stellar equator in kG, $R_2$ is the stellar radius in units of 2 $R_{\odot}$, $M_{0.5}$ is the stellar mass in units of 0.5 M$_{\odot}$ and $\dot{M}_{-8}$ is the mass-accretion rate in units of $10^{-8}$ \umdot{}.

Using the stellar parameters of CI Tau reported by \citet{2020MNRAS.491.5660D}, for a magnetic field of 0.85 kG\footnote{We use the polar magnetic field value of 1.7 kG divided by two to retrieve the value at the equator.}, a stellar radius of $2.0\pm0.3$ R$_{\odot}$, a mass of $0.90\pm0.02$ M$_{\odot}$ \citep{Simon19}, and the mass-accretion rate derived from Eq. \ref{eq:macc}, we compute a truncation radius \rtr{} $= 3.6\pm1.5\,R_\star$, in agreement with the interferometric half-flux radius derived above for the \brg{} line emitting region. We therefore conclude that most of the \brg{} emission originates from the magnetospheric accretion region.

\begin{figure*}[h!]
    \centering
    \includegraphics[width=0.9\textwidth]{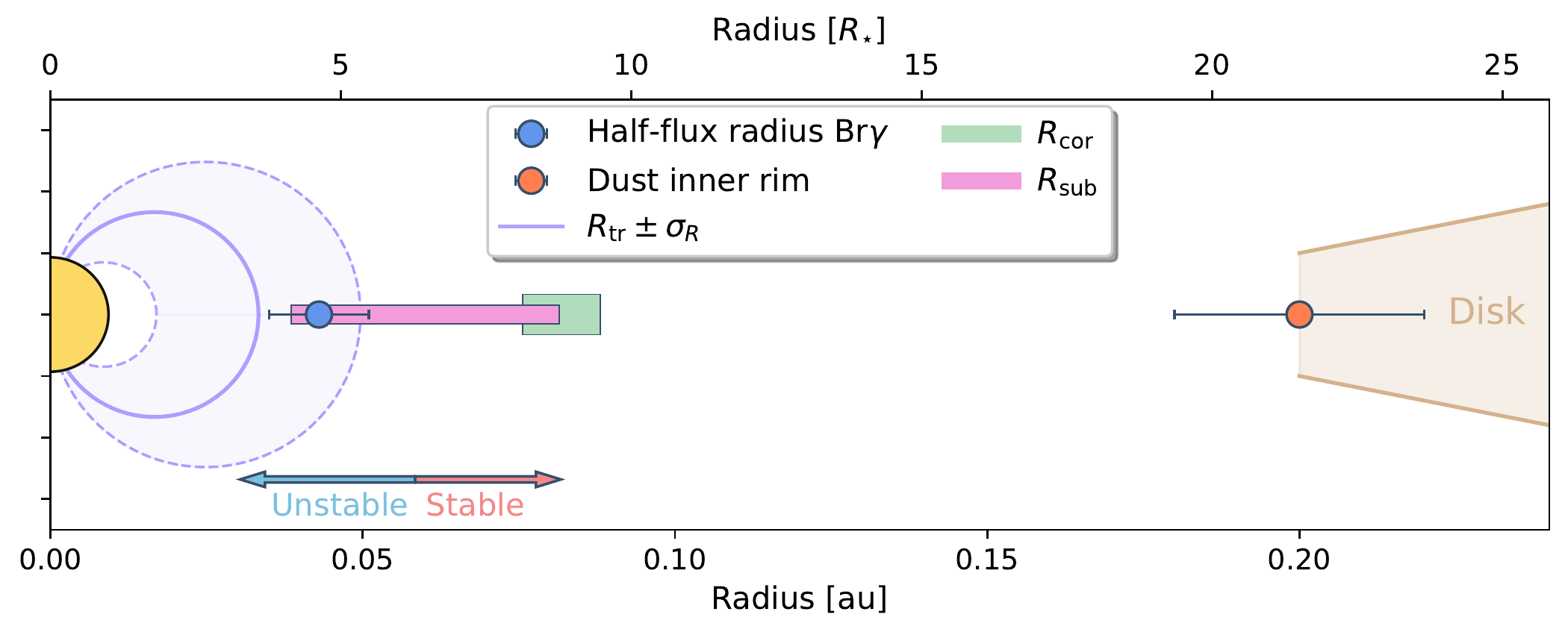}
    \caption{\label{fig:resum_gravity} A schematic view of the innermost region of the CI Tau system. The sizes and their uncertainties derived from the GRAVITY observations are represented for the \brg{} emitting region (\textit{blue circle}) and for the K-band continuum (\textit{orange circle}). The purple circle (and shaded area) depicts the truncation radius and its uncertainty derived from the GRAVITY \brg{} emission line (Sect. \ref{sec:mass_acc_measure}). Additional characteristic scales associated with YSO are depicted: the sublimation radius (\rsub{}, {\it pink line}) for a range of absorption $Q_R$ from 1 to 4 and the co-rotation radius (\rco{}, {\it green line}). The stable-unstable magnetospheric accretion regimes are indicated  with a boundary around $70\%$ of the \rco{} \citep{blinova16}.} 
\end{figure*}

\section{Discussion}

The high spatial and spectral resolution of GRAVITY allows us to detect and characterise the inner region of the CI Tau system with an unprecedented precision. Figure \ref{fig:resum_gravity} illustrates the characteristic sizes of the system. In this section, we discuss how the GRAVITY results shed light on the global structure of the inner system.

\subsection{The inner dust cavity}

The continuum analysis of the two epochs of observation yields an inner dusty rim  located between 14 and 21 $R_{\star}$ from the central star. This direct measurement appears to be significantly higher than the estimate of the dust sublimation radius. The K-band emission of T Tauri stars is supposed to be dominated by the directly irradiated front of the dusty disk rim \citep{2010ARA&A..48..205D}. For a given stellar luminosity, we assess the radius corresponding to the thermal equilibrium of the dust grains, to remain under the sublimation temperature at 1500 K for silicates. We used the relation from \citet{2002ApJ...579..694M} to determine the sublimation radius (\rsub{}) in au:
\begin{equation}
 R_{\mathrm{sub}} = 1.1\sqrt{Q_R}\sqrt{\frac{L_\star}{1000\,L_\odot}}\left(\frac{1500}{T_{\mathrm{sub}}}\right)^2,
\end{equation}
with $Q_R$ the absorption efficiency ratio of the dust between incident and reemitted field, and $T_{\mathrm{sub}}$ the sublimation temperature. \citet{2002ApJ...579..694M} assess that the absorption efficiency $Q_R$ depends on the dust properties and the effective temperature of the central star. For an effective temperature of 4200 K \citep{2020MNRAS.491.5660D} and a typical grain size distribution ranging from 0.03 to 1 $\mu$m, $Q_R$ ranges from 1 to 4. For a stellar luminosity of 1.26 $L_\odot$ \citep{2020MNRAS.491.5660D}, the 1500~K sublimation radius ranges from 0.04 to 0.08~au, i.e., 4.3 to 8.6 $R_\star$. The inner disk rim location we derive is therefore at least twice farther than the sublimation radius (see Fig. \ref{fig:resum_gravity}), when considering a sublimation temperature of 1500~K.

One potential explanation for an extended inner dust cavity is the presence of an hypothetical close-in planet. CI Tau is so far the only Class II pre-main sequence star claimed to host a hot Jupiter, CI Tau b, with a mass of $\sim$11.3 Jupiter mass \citep{2016ApJ...826..206J, 2019ApJ...878L..37F}. If such a planet exists, it could significantly affect the inner region of the disk. \citet{2021ApJ...921L..34M} demonstrated that a massive candidate planet orbiting at 0.08~au leads to the formation of an inner gap ranging from 0.1 to 0.2~au depending on the eccentricity of the planet, fully compatible with our observation.



\subsection{The inner and outer disk misalignment}

Young stellar objects such as the CI Tau system harbour a large outer disk structure. \citet{2018ApJ...866L...6C} retrieve the geometrical properties of CI Tau's outer disk on a scale from 1 to 100~au using the Atacama Large Millimeter/Submillimeter Array (ALMA). The outer disk consists of multiple rings seen at an inclination of $i_{\mathrm{out}}=50\degree$ and a position angle of $PA_{\mathrm{out}}\simeq11\degree$ from North to East. In comparison, the inner disk orientation we derive from GRAVITY features $i_{\mathrm{in}}\simeq70\degree$ and $PA_{\mathrm{in}}=148\degree$. The two disks thus appear significantly misaligned. Such a misalignment has been recently reported on a few targets among a large sample of YSO \citep{2022A&A...658A.183B}. 

Following \citet{2017A&A...604L..10M, 2022A&A...658A.183B}, we can thus measure the misalignment angle between the inner and outer disks as:
\begin{equation}
\begin{split}
\Delta\theta(i_{\mathrm{in}}, PA_{\mathrm{in}}, i_{\mathrm{out}}, PA_{\mathrm{out}})
&=\arccos[\sin(i_{\mathrm{in}})\sin(i_{\mathrm{out}})\\&\times\,\cos(PA_{\mathrm{in}}-PA_{\mathrm{out}})\\
&+\cos(i_{\mathrm{in}})\cos(i_{\mathrm{out}})]
\end{split}
\end{equation}

The misalignment angle $\Delta\theta$ corresponds to the angle between the two normal vectors defined by the planes of the inner and outer disk. Additionally, we do not know which side of the inner disk is closest to the observer. Two misalignment angles can therefore be calculated, namely $\Delta\theta_1\sim109$ or $\Delta\theta_2\sim42\degree$ for CI~Tau. In both cases, the inner and outer disks appear to be significantly misaligned. While such a significant misalignment may induce a shadow projected onto the outer disk \citep{2022A&A...658A.183B}, such a shadow is not detected in scattered light images of CI Tau's disk \citep{2022A&A...658A.137G}. 

Various physical processes can induce a substantial misalignment between the inner and the outer disks. Gravitational torques caused by the presence of low-mass \citep{2018MNRAS.475.3201A} or high-mass \citep{2013MNRAS.431.1320X} planets can force the precession of the inner disk, and physically disconnect it from the outer disk. For the massive case (> 1 $M_{\mathrm{Jup}}$), if the companion's angular momentum is significantly greater than the disk one, the inner disk can gain a warped inner structure with an inclination of up to $\simeq20\degree$ relative to the outer part. Recent 3-d simulations reinforce this assumption for planets massive enough to carve gaps \citep{2018MNRAS.481...20N}. Inner-outer disk misalignments are not only observed as a consequence of massive companions. Differential angular momentum across the disk can induce a tilt between the spin vectors of the various components (star, inner and outer disks) \citep[][]{2022ApJ...931...42E}. The magnetic star-disk interaction can also warp the close-in region and be responsible for an inclined inner disk, up to 40° inclination, with respect to the stellar-spin axis \citep{2021MNRAS.506..372R}. Finally, an external infall of gaseous material could affect the outer disk region and induce a misalignment \citep{2021A&A...656A.161K}. A detailed review of the misalignment processes and shadowing effects is provided in \citet{2022arXiv220309991B}. 

\subsection{The magnetospheric accretion region}

From the exquisite precision of the differential visibilities achievable with GRAVITY, we are able to spatially resolve the characteristic size of the \brg{} line emitting region. With a half-flux radius of $0.045^{+0.008}_{-0.009}$~au, a large fraction of the \brg{} line emission appears to originate from a region extending over $4.8^{+0.8}_{-1.0}$ $R_\star$ around the star. 

A quantitative comparison between the \brg{} half-flux radius and the co-rotation radius can be used as a simple criterion to determine the physical origin of the observed \brg{} line emission. If the \brg{} emission appears as a compact source, smaller than the co-rotation radius, the origin is consistent with the magnetospheric accretion scenario. In contrast, if the \brg{} emission is significantly larger than the co-rotation radius, other mechanisms such as disk winds or outflows are likely to contribute to the observed \brg{} profile \citep{Gravity20}. The co-rotation radius is defined as the one where the angular velocity of the rotating disk matches the angular velocity of the star:
\begin{equation}
    R_{\mathrm{cor}}=(GM_\star)^{1/3}(P_\mathrm{rot}/2\pi)^{2/3}
\end{equation}
For a rotational period $P_{\mathrm{rot}}=9.00\pm0.05$ days \citep{2020MNRAS.491.5660D} and a mass of $0.90\pm0.02$ M$_{\odot}$ \citep{Simon19}, we compute a co-rotation radius \rco{} $=8.8\pm1.3\,R_\star$.

We find here that the \brg{} half-flux radius is significantly smaller than the co-rotation radius, which argues in favour of most of the line flux arising from the magnetospheric accretion process. Furthermore, based on spectro-polarimetric magnetic field measurements, \citet{2020MNRAS.491.5660D} estimated a range of values between 3.7 and 6.3 $R{_\star}$ for the magnetospheric truncation radius of CI~Tau, which is consistent with our GRAVITY measurement of $3.6\pm1.5~R{_\star}$. We caution, however, that the interferometric \brg{} half-flux radius derived from a 2-d Gaussian model may underestimate the full extent of the magnetospheric accretion region \citep{Tessore23}.


From our truncation radius estimate (Sect. \ref{sec:mass_acc_measure}), we derive a ratio of \rtr{}/\rco{} = $0.41\pm0.18$ and the system will likely be in an unstable accretion regime \citep[\rtr{}/\rco{}$\lesssim0.7$, ][]{blinova16}. In magnetic star-disk interactions, unstable accretion is the outcome of an interchange instability where the gas penetrates the stellar magnetosphere through equatorial tongues \citep{Romanova08} in addition to the stable funnel flows (i.e., stable accretion). Such accretion tongues are expected to deposit matter at random places on the stellar surface, usually close to the stellar equator, a feature that can possibly explain
the stochastic photometric behaviour of the system known as a burster \citep{roggero21, cody22}. 


\section{Conclusion}

We have used the VLTI/GRAVITY instrument to probe the innermost scales of the young system CI Tau. Investigating the K-band spectral domain at high spectral resolution allows us to study the system in the continuum to probe dust emission and within the \brg{} line to trace gas emission simultaneously. Below, we summarise our major results.

(i) From the continuum analysis, we report the detection of a highly inclined resolved inner disk, whose inner edge is located at a distance of $21\pm2\,R_\star$ from the central star. The measured inner rim position seems to be significantly farther than the theoretical sublimation radius (4-8 $R_\star$ for a typical sublimation temperature of silicates of 1500~K), a result which might support the presence of a close-in massive planetary companion.

(ii) The inner disk exhibits a strong misalignment relative to the outer disk seen at submillimeter wavelengths with ALMA. Such a misalignment could be induced by magnetic warping or by gravitational torques induced by a close-in massive companion.

(iii) We constrained the half-flux radius of the \brg{} emitting region to be at a distance of 4.8 $R_\star$ from the central star, which is consistent with the magnetospheric accretion paradigm. The \brg{} size is significantly smaller than the co-rotation radius, which leads to an unstable accretion regime, presumably at the origin of the stochastic photometric variability of the system.

The interferometric precision achievable today with GRAVITY at the VLTI allows us to characterise the inner scales of the CI Tau system with an unprecedented sensitivity. Given the high variability of this system, a temporal follow-up represents the most promising opportunity to investigate the dynamics of the star-disk interaction process, and to ascertain the origin of the \brg{} emission. This work represents a first step to understand the star-planets-disk interactions occurring on sub-au scales in young stellar objects.

\newpage
\begin{acknowledgements}

We acknowledge support from the European Research Council (ERC) under the European Union’s Horizon 2020 research and innovation programme (grant agreement No 742095; {\it SPIDI}: Star-Planets-Inner Disk-Interactions, \url{http://www.spidi-eu.org}). We thank A. Sousa for providing the infrared veiling measurements. We thank M. Benisty for the fruitful discussion about the disk misalignment and for confirming the misalignment values. We thank A. Wojtczak for the pure line derivation and the benchmark of our algorithms. We thank in particular the SPIDI crew for providing ideas and triggering discussions on the accretion phenomenon (B. Tessore, R. Manick). A.C.G. has been supported by PRIN-INAF MAIN-STREAM 2017 “Protoplanetary disks seen through the eyes of new-generation instruments” and PRIN-INAF 2019 “Spectroscopically tracing the disk dispersal evolution (STRADE)”. This work has made use of data from the European Space Agency (ESA) mission Gaia (\url{https://www.cosmos.esa.int/gaia}), processed by the Gaia Data Processing and Analysis Consortium (DPAC, \url{https://www.cos-mos.esa.int/web/gaia/dpac/consortium}). This research made use of NASA's Astrophysics Data System; \textsc{SciPy} \citep{scipy}; \textsc{NumPy} \citep{numpy}; \textsc{matplotlib} \citep{matplotlib}; and Astropy, a community-developed core Python package for Astronomy \citep{astropy}. This research has made use of the Jean-Marie Mariotti Center \texttt{LITpro}\footnote{LITpro software available at \url{http://www.jmmc.fr/litpro}}, \texttt{Aspro2}\footnote{Available at \url{http://www.jmmc.fr/aspro2}} and \texttt{SearchCal}\footnote{Available at \href{https://www.jmmc.fr/english/tools/proposal-preparation/search-cal/}{\url{https://www.jmmc.fr/searchcal}}} services, co-developped by CRAL, IPAG and LAGRANGE,.
 
\end{acknowledgements}

\bibliographystyle{aa}	
\bibliography{citau_gravity} 

\appendix

\section{Data-model comparison}
\label{app:fit_results}

\begin{figure}[h]
    \centering
    \includegraphics[width=0.88\columnwidth]{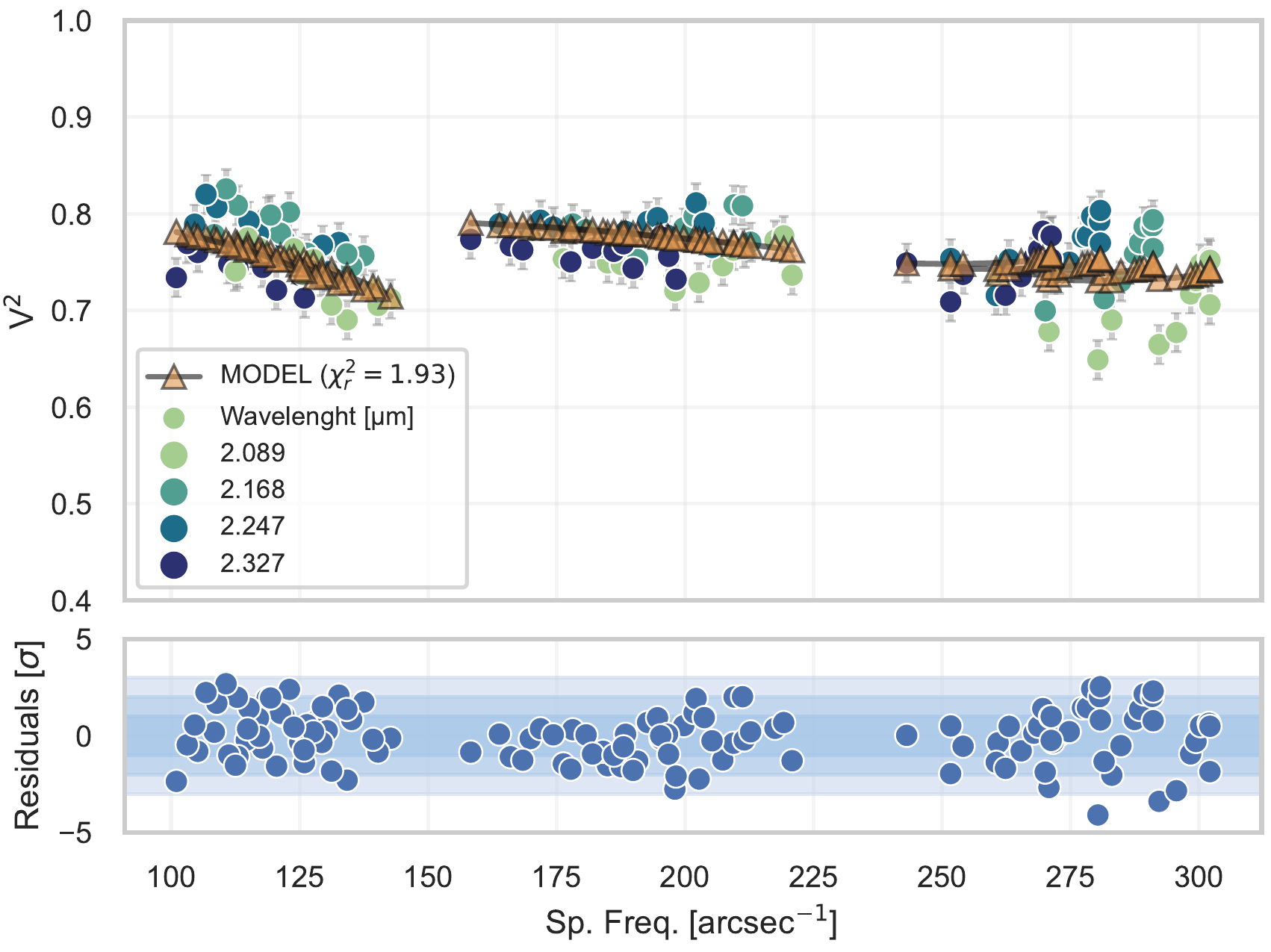}
    \caption{\label{fig:fit_V2_2021} Comparison of the best-fitted model with the squared visibilities for epoch 1. The colours illustrate the wavelengths of the four spectral channels used to fit the K-band continuum. The model is presented in orange. The lower panel shows the residuals compared to 1-$\sigma$, 2-$\sigma$ and 3-$\sigma$ relative errors.}  
\end{figure}

\begin{figure}[h!]
    \centering
    \includegraphics[width=0.88\columnwidth]{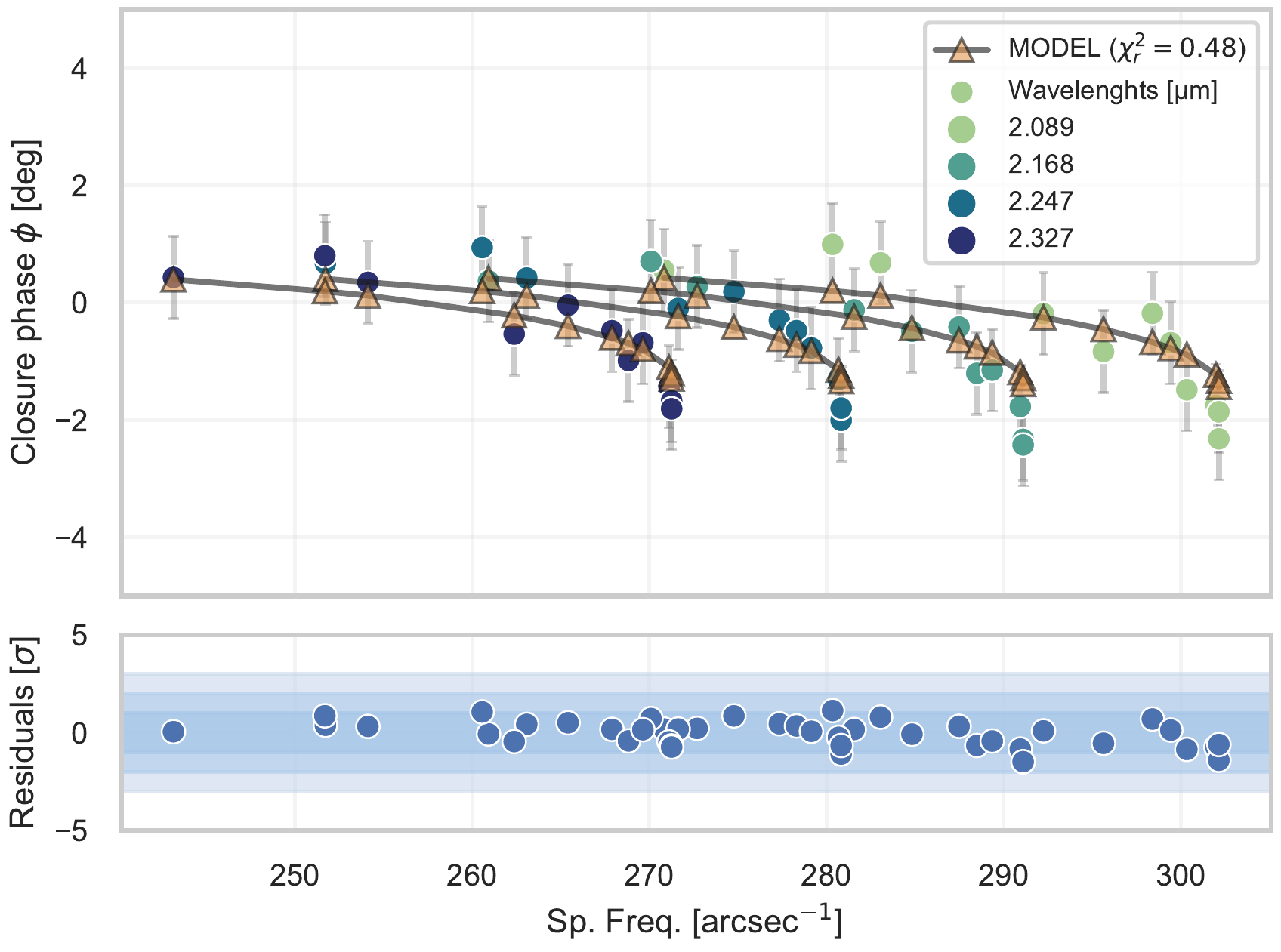}
    \caption{\label{fig:fit_CP_2021} Comparison of the best-fitted model with the closure phases for epoch 1. Same description as in Fig. \ref{fig:fit_V2_2021}. Note that we present only one closure phase triangle due to the missing UT2 during the observations.}  
\end{figure}

\begin{figure}[h!]
    \centering
    \includegraphics[width=0.88\columnwidth]{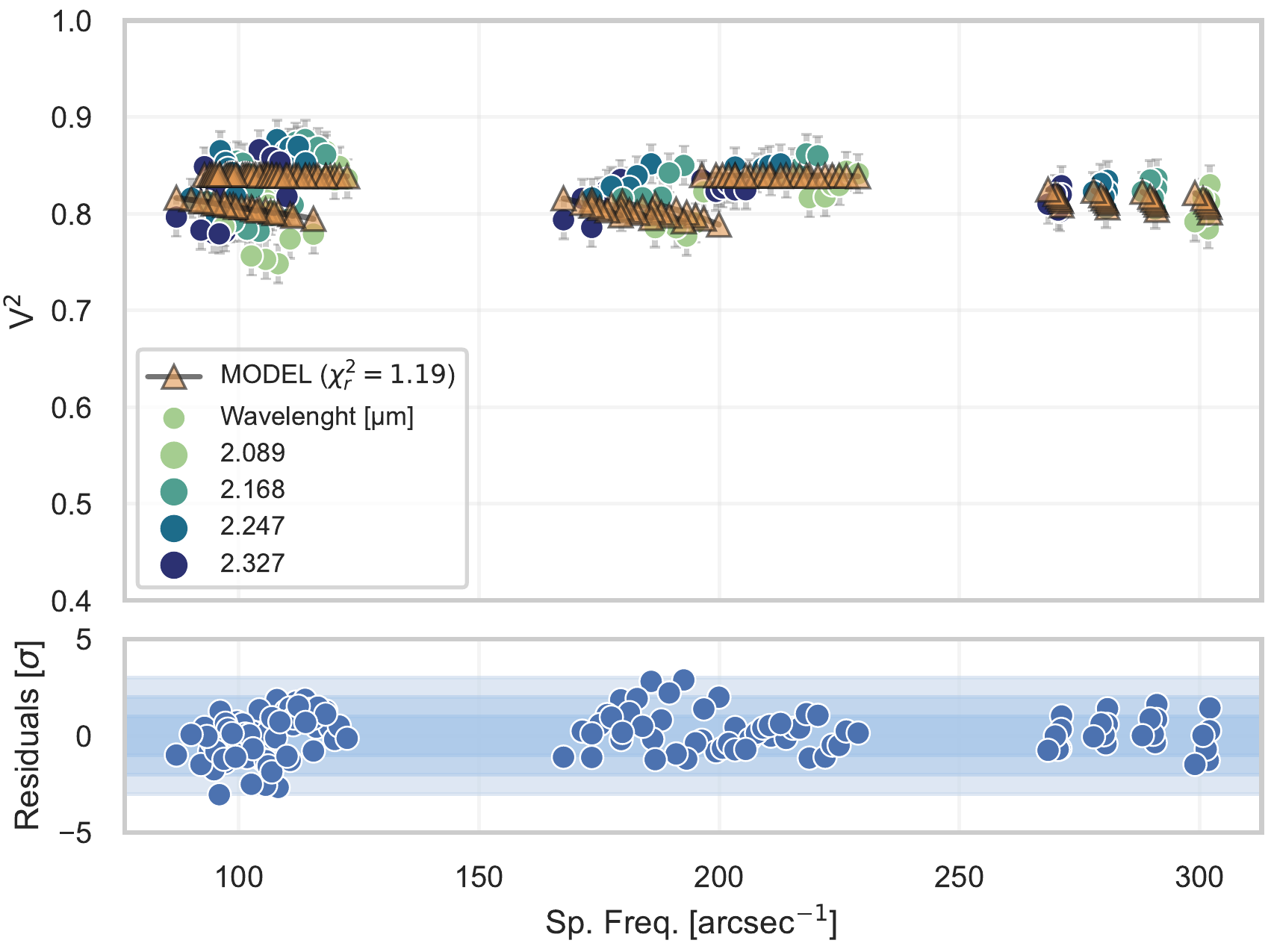}
    \caption{\label{fig:fit_V2_2022} Same description as in Fig. \ref{fig:fit_V2_2021} for the 2022 dataset.}  
\end{figure}

\begin{figure}[h!]
    \centering
    \includegraphics[width=0.88\columnwidth]{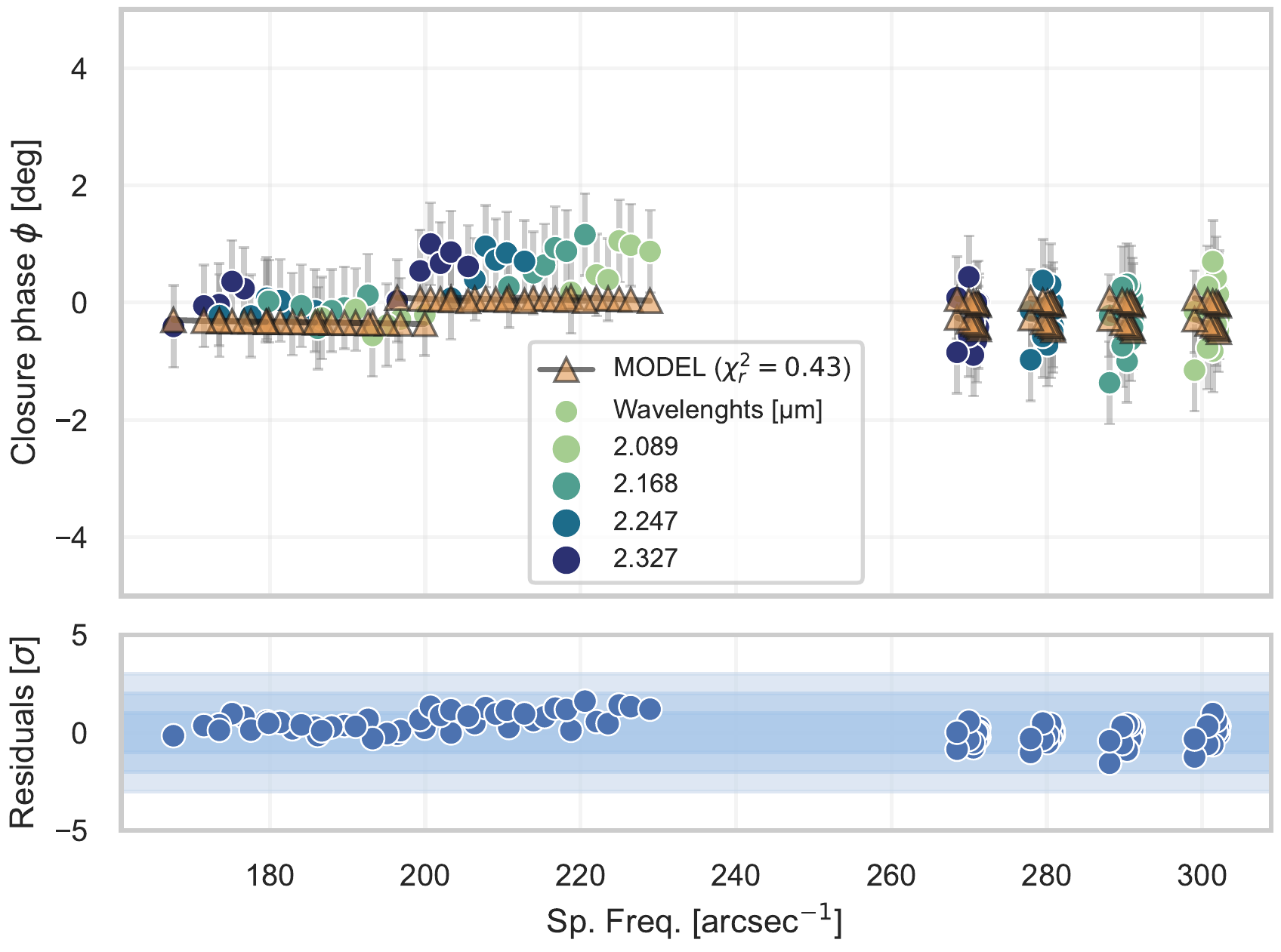}
    \caption{\label{fig:fit_CP_2022} Same description as in Fig. \ref{fig:fit_CP_2021} for the 2022 dataset.}  
\end{figure}

In this appendix, we present the results and data-model comparisons of the MCMC posterior determination of the K-band continuum geometrical model for the two epochs of observations. Figures \ref{fig:fit_V2_2021}, \ref{fig:fit_CP_2021}, \ref{fig:fit_V2_2022} and \ref{fig:fit_CP_2022} show the comparison between the best fitted-model and the data for the squared visibilities and closure phases, respectively. The visibilities indicate a resolved structure ($V^2 < 0.8$), while the non-zero closure phase ($< 2\degree$) points to an asymmetric environment. The small tilt of the shortest baseline compared to the middle-range baseline suggests an inclined object, as retrieved by our image model (Fig. \ref{fig:model_image}).

\section{$\chi^2$-search uncertainty refinement}
\label{app:chi2_fit_incl}

\begin{figure}[h!]
    \centering
    \includegraphics[width=0.92\columnwidth]{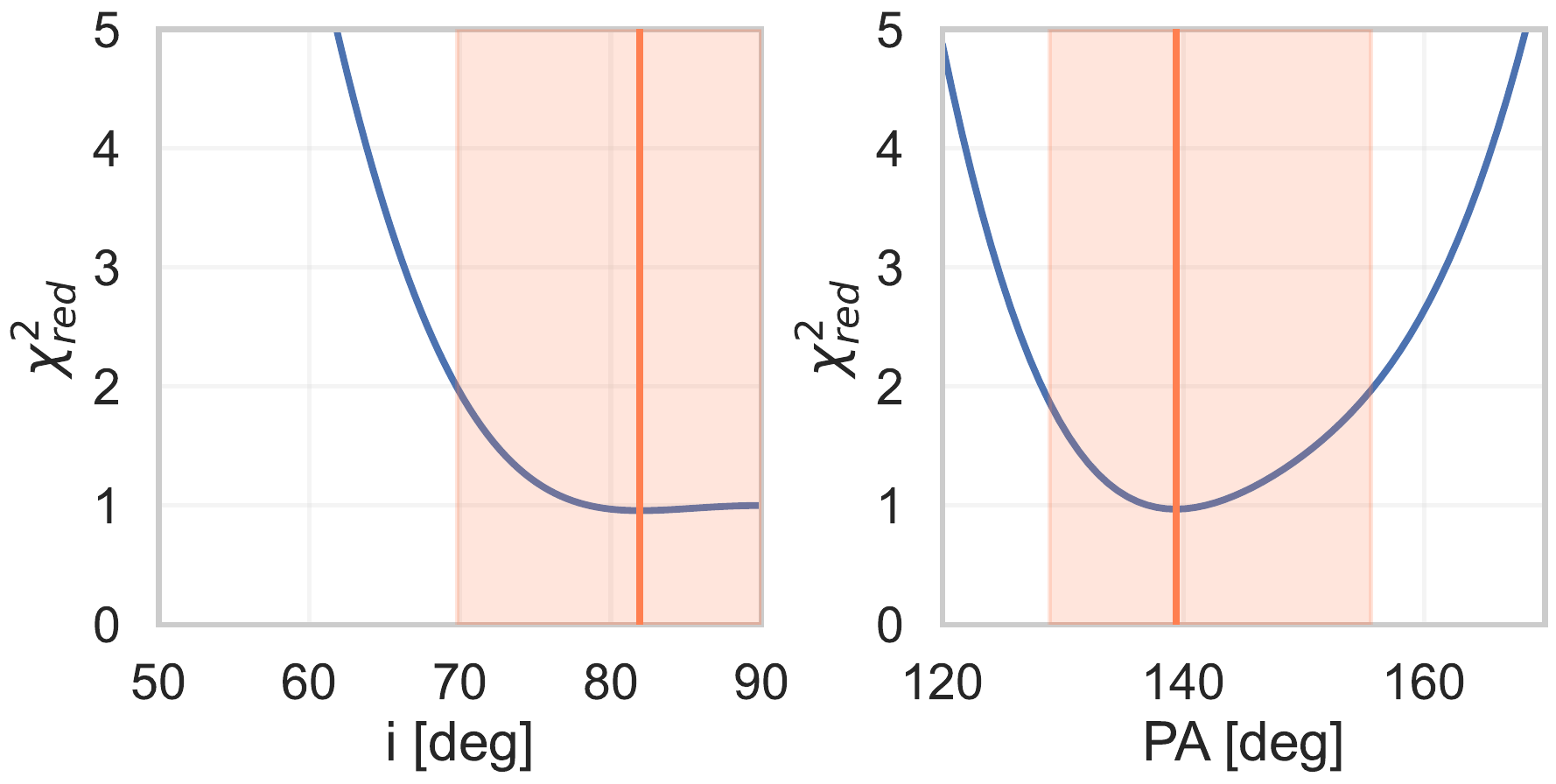}
    \caption{\label{fig:chi2_orient_limit} $\chi^2$-curves search around the MCMC-posterior  results of the second epoch 2022. \textbf{Left:} inclination between 50 and $90\degree$. \textbf{Right:} Position angle between 120 and $170\degree$. The orange vertical line denotes the posterior mean value. The orange shade area represents the 1-$\sigma$ uncertainty corresponding to $\chi^2_{red, min}+1$.}
\end{figure}

\begin{figure}[h!]
    \centering
    \includegraphics[width=0.93\columnwidth]{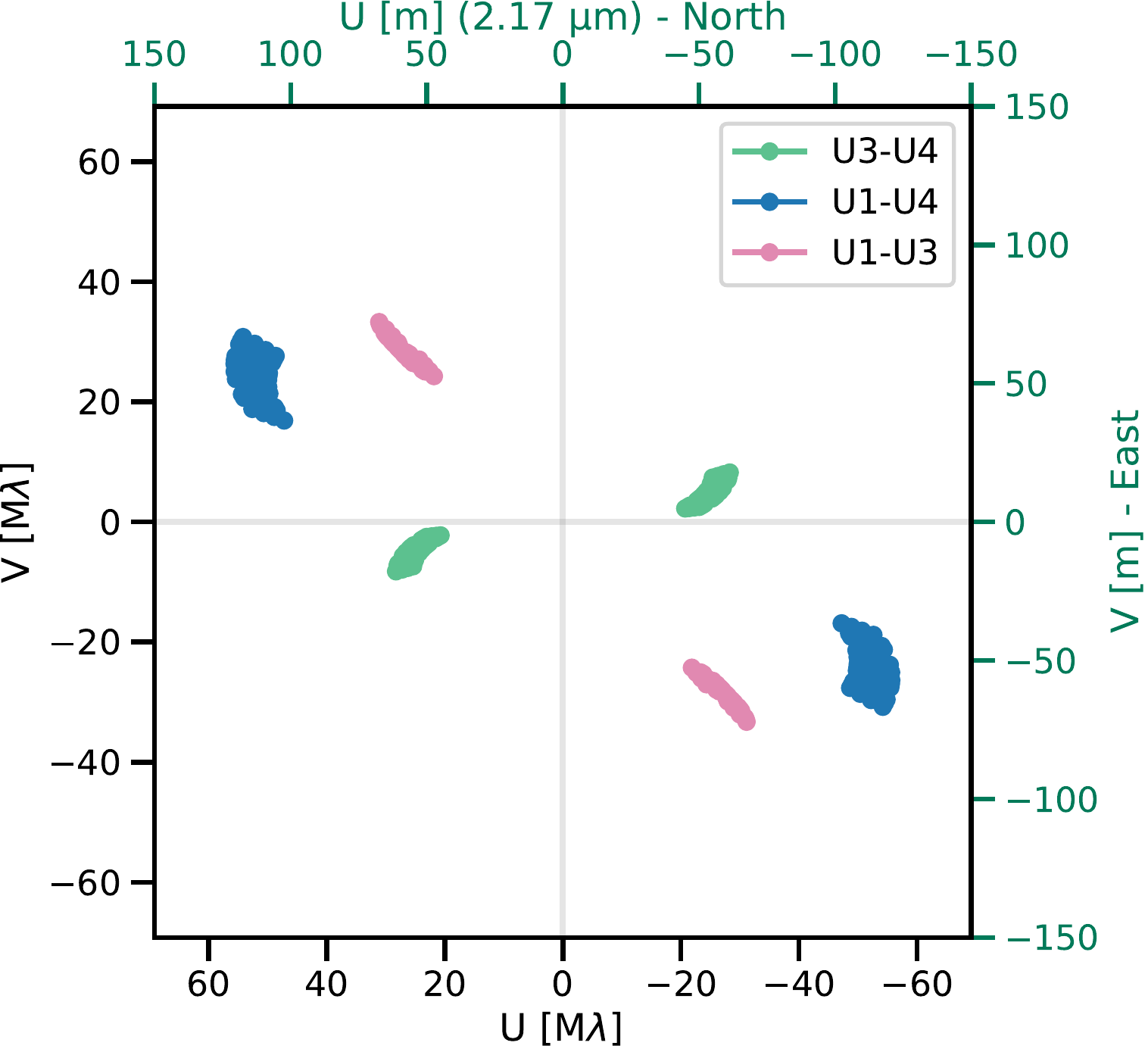}
    \caption{\label{fig:uv_plan_2021} Fourier coverage obtained for the 2021 epoch.}
\end{figure}

\begin{figure}[h!]
    \centering
    \includegraphics[width=0.93\columnwidth]{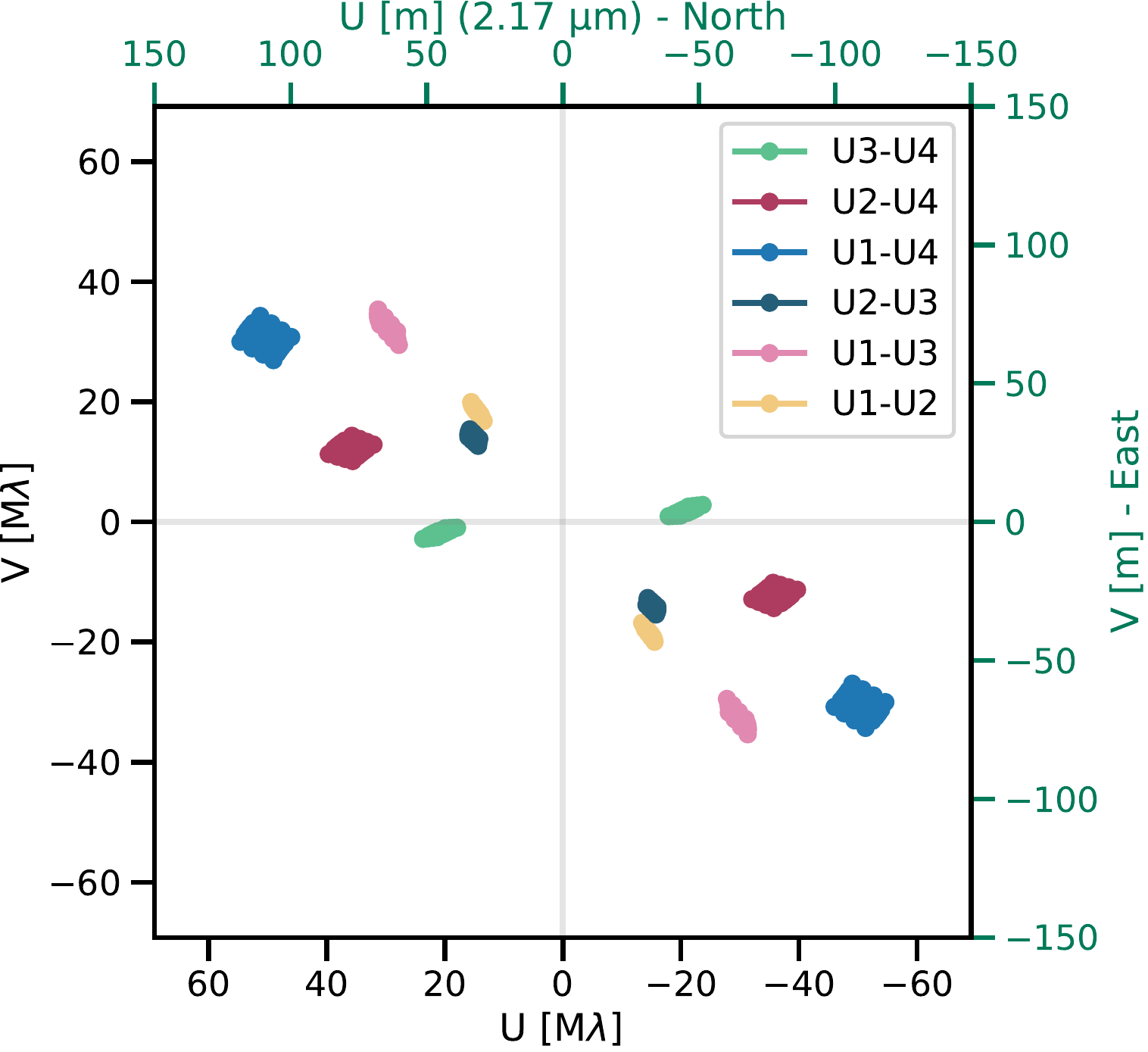}
    \caption{\label{fig:uv_plan_2022} Fourier coverage obtained for the 2022 epoch.}
\end{figure}

Figure \ref{fig:chi2_orient_limit} shows the $\chi^2$ data-comparison values for a range of inclination ($i$) and position angle ($PA$). This conservative method allows us to give a lower limit on the inclination parameters, not well constrained by the MCMC-posterior estimation due to the limited range of spatial frequencies covered during the second epoch (30 vs. 75 arcsec$^{-1}$ in 2021). The two u-v coverages are presented in Figure \ref{fig:uv_plan_2021} and \ref{fig:uv_plan_2022}. When using the reduced $\chi^2$ as a limiting factor, we consider fully correlated error bars on the data. This method tends to overestimate the errors associated with the physical parameters we aim to constrain. In our case, the system orientation is consistent with a high inclination ($i>70\degree$) at a position angle in agreement with our 2021 estimate.

\newpage
\section{MCMC posterior distribution}
\label{app:mcmc_distr}

\begin{figure*}
    \centering
    \includegraphics[width=0.98\textwidth]{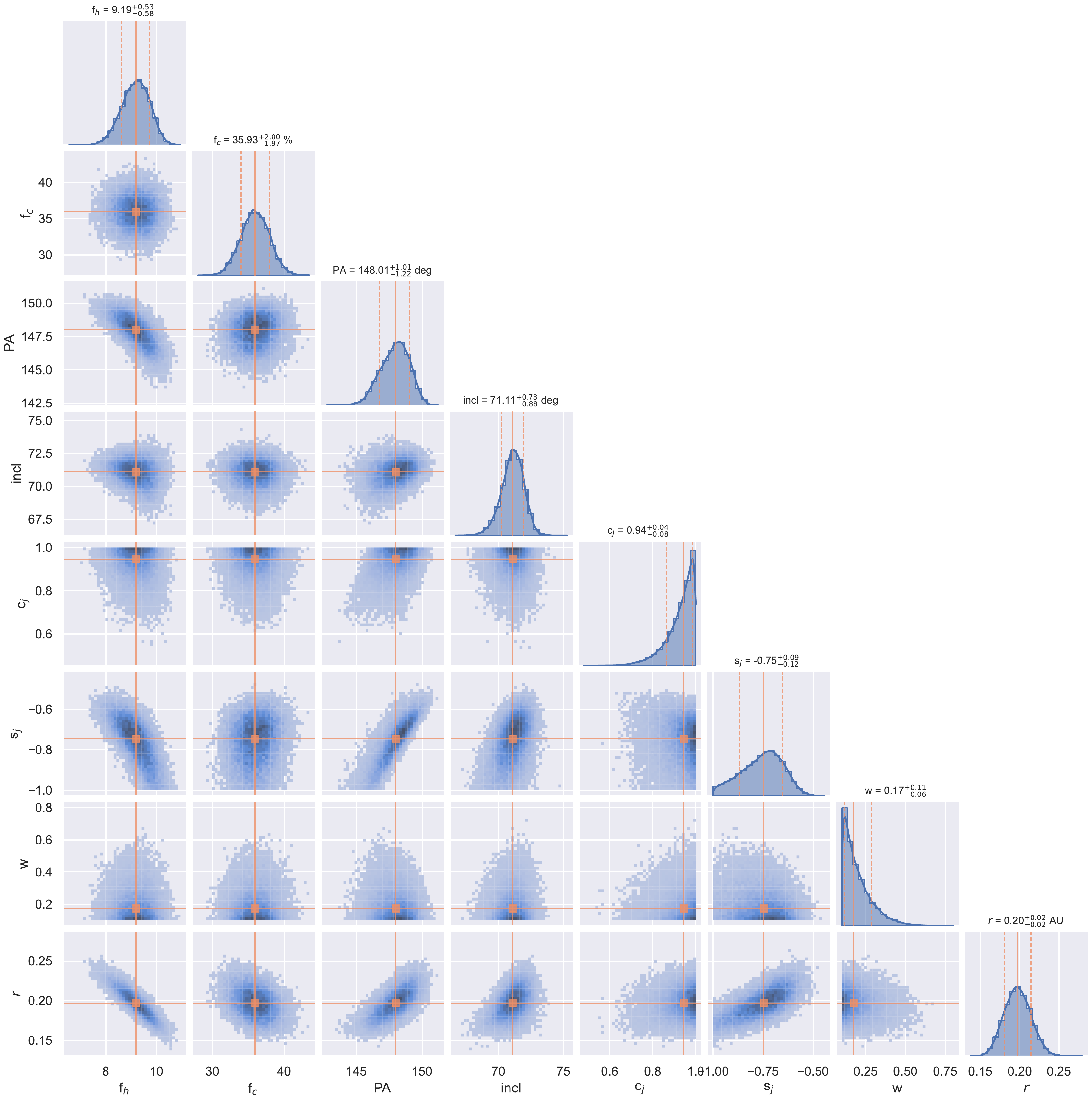}
    \caption{\label{fig:mcmc_distribution} Final distribution of parameter values deduced from the MCMC computation for the 2021 data. The standard percentiles are used to derive the mean values and the uncertainties.}  
\end{figure*}

\begin{figure*}
    \centering
    \includegraphics[width=0.98\textwidth]{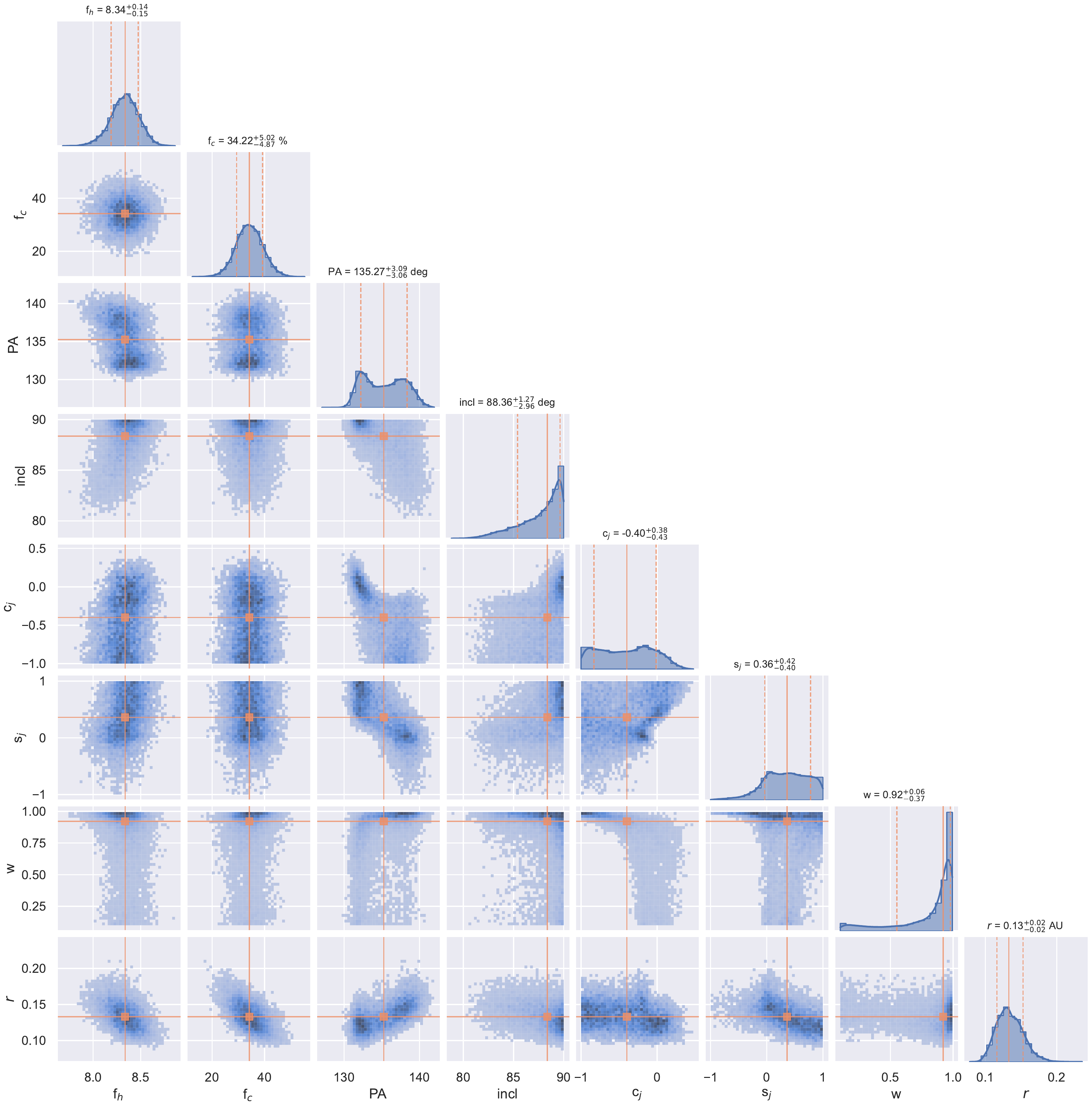}
    \caption{\label{fig:mcmc_distribution2} Same as Fig. \ref{fig:mcmc_distribution} for the second epoch in 2022.}
\end{figure*}

Figure \ref{fig:mcmc_distribution} and \ref{fig:mcmc_distribution2}  show the final posterior distribution of walkers corresponding to 200,000 individual iterations. The MCMC-fit converges toward a unique model of an inclined thin disk for the first epoch (2021). The parameters estimations are presented in Table \ref{tab:results_fit_cont}. In 2022, the posterior distribution does not allow us to determine the system inclination or the width-radius ratio ($w$) unambiguously. The one-hour time range observation available in 2022 does not offer sufficient rotation of the u-v coverage to derive these two parameters accurately, contrarily to the three-hour range time achieved in 2021. The ratio $w$ close to one indicates that we are not able to constrain the gap size reported in 2021. Additionally, the modulation parameters $c_j$ and $s_j$ are compatible with zero. Nevertheless, we are able to deliver strong constraints on the half-flux radius of the disk, the position angle and components contributions (see Table \ref{tab:results_fit_cont}).

\end{document}